\documentstyle[graphicx,aps,multicol]{revtex}
\draft

\begin{document}

\title{Model for pairing phase transition in atomic nuclei}
\author{A.~Schiller\footnote{Electronic address: 
Andreas.Schiller@llnl.gov}}
\address{Lawrence Livermore National Laboratory, L-414, 7000 East Avenue, 
Livermore CA-94551}
\author{M.~Guttormsen, M.~Hjorth-Jensen, J.~Rekstad, and S.~Siem}
\address{Department of Physics, University of Oslo, N-0316 Oslo, Norway}
\maketitle

\begin{abstract}
A model is developed which allows the investigation and classification of the 
pairing phase transition in atomic nuclei. The regions of the parameter space 
are discussed for which a pairing phase transition can be observed. The model 
parameters include: number of particles, attenuation of pairing correlations 
with increasing seniority, single particle level spacing, and pairing gap 
parameter. It is argued that for nuclear models, the pairing phase transition 
has to be separated in temperature from the signal of the exhaustion of the 
finite model space.
\end{abstract}


\begin{multicols}{2}

\section{Introduction}

Phase transitions in small systems are an important research topic. One of the 
most interesting problems in this context is probably the question of the 
existence and classification of a possible phase transition from a hadronic 
phase to a quark-gluon plasma in high energy physics. The answer to this 
question has far-reaching consequences into many other fields of research like,
e.g., cosmology, since it has been argued that hadronization of the quark-gluon
plasma should be a first-order phase transition in order to allow for possible 
supercooling and consequently the emergence of large scale inhomogeneities in 
the cosmos within the inflationary big-bang model \cite{Da01}. 

In nuclear physics, two different phase transitions have been discussed in the 
literature. A first-order phase transition has been reported in the 
multifragmentation of nuclei \cite{AG00}, thought to be the analogous 
phenomenon in a finite system to a liquid-gas phase transition in the 
thermodynamical limit. A pivotal role in these studies is played by the 
presence of a convex intruder in the microcanonical entropy curve 
\cite{Gr97,GV00}. This leads to a negative branch of the microcanonical heat 
capacity which is used as an indicator of a first-order phase transitions in 
small systems. Negative heat capacities have indeed been observed in the 
multifragmentation of atomic nuclei, though the heat capacity curve has not 
been derived directly from the caloric curve, but by means of energy 
fluctuations \cite{AG00,CD00}. Another finding of a negative branch of the heat
capacity curve has been in sodium clusters of 147 atoms \cite{SK01}, indicating
a possible first-order phase transition. On the other hand, it is not clear 
whether the observed negative heat capacities are simply due to the changing 
volume of the system under study that is progressively evaporating particles 
\cite{ME01}. It has also been suggested that a negative heat capacity can be 
observed when studying a system in a metastable state \cite{TC02}. In general, 
great care should be taken in the proper extraction of temperatures and other 
thermodynamical quantities of a multifragmenting system \cite{PI96,DM00}.

A further difficulty in characterizing the order of a phase transition in a 
finite system using negative heat capacities arises from the fact that there
are no counterparts in the thermodynamical limit. For this reason, no simple 
connection with the Ehrenfest classification of phase transitions can be made. 
Therefore, it is not clear that a phenomenon involving negative heat capacities
in small systems is analogous to a first-order phase transition in the 
thermodynamical limit. Another way to classify phase transitions for finite 
systems has been proposed within the canonical ensemble. This classification 
scheme is based on the distribution of zeros (DOZ) of the canonical partition 
function in the complex temperature plane \cite{BM00} and has successfully been
applied to a model of multifragmentation \cite{MB01}. This classification 
scheme reduces to the Ehrenfest classification in the thermodynamical limit and
is applicable for first, second, and higher order phase transitions.

The second phase transition discussed for atomic nuclei has been anticipated 
for the transition from a phase with strong pairing correlations to a phase 
with weak pairing correlations \cite{SY63}. Early schematic calculations have 
shown that pairing correlations can be quenched by temperature as well as by 
the Coriolis force in rapidly rotating nuclei \cite{Go81a,Go81b,TS80,TS82}. 
This makes the quenching of pairing correlations in atomic nuclei very similar 
to the breakdown of superfluidity in $^3$He (due to rapid rotation and/or 
temperature) or of superconductivity (due to external magnetic fields and/or 
temperature). Recently, structures in the heat capacity curve related to the 
quenching of pairing correlations have been obtained within the relativistic 
mean field theory \cite{AT00,AT01}, the finite-temperature random phase 
approximation (RPA) \cite{Ng90}, the finite-temperature Hartree-Fock-Bogoliubov
theory \cite{ER00}, and the shell-model Monte Carlo (SMMC) approach 
\cite{DK95a,NA97,RH98,WK00,LA01}. An S-shaped structure in the heat capacity 
curve could also be observed experimentally \cite{SB01} and has been 
interpreted as a fingerprint of a second-order phase transition from a phase 
with strong pairing correlations to a phase with weak pairing correlations. 
Indeed, the analogy of the quenching of pairing correlations in atomic nuclei 
with the breakdown of superfluidity in $^3$He and the breakdown of 
superconductivity suggests a second order phase transition and a schematic 
calculation might support this assumption \cite{BD01}. Interestingly, similar 
structures of the heat capacity curve as observed for atomic nuclei in 
\cite{SB01} have been seen in small metallic grains undergoing a second-order 
phase transition from a superconductive to a normal conductive phase 
\cite{LA93}, thereby supporting the analogous findings for atomic nuclei. On 
the other hand, breaking of nucleon pairs has been experimentally shown to 
cause a series of convex intruders in the microcanonical entropy curve of 
deformed rare earth nuclei \cite{MB99,MG01}, leading to several negative 
branches of the microcanonical heat capacity. This finding might, in analogy to
the discussion of nuclear multifragmentation, be taken as an indicator of 
several first-order phase transitions. Interpreting the quenching of pairing 
correlations as a series of first-order phase transitions seems, however, 
physically unattractive and might rather suggest that a negative branch of the 
microcanonical heat capacity curve does not necessarily indicate the analogous 
phenomenon in a small system to a first-order phase transition in the 
thermodynamical limit. A similar conclusion has been drawn by the authors of 
Ref.~\cite{BD01}.

We will in this work develop a model for the atomic nucleus which allows us to 
investigate the occurrence and classification of the pairing phase transition 
in atomic nuclei as a function of particle number and other parameters within 
the model. In order to do so, we will apply the classification scheme of 
Borrmann \sl et al.\ \rm \cite{BM00,MB01} to the pairing phase transition. 

\section{Statistical ensembles}

The proposed nuclear model is formulated within the canonical ensemble theory 
although the microcanonical ensemble theory might, at a first glance, be the 
more appropriate ensemble for describing the nucleus below the particle 
threshold. At low excitation energies, the nucleus can be regarded as a closed 
system with respect to energy and particle exchange. However, the 
microcanonical ensemble theory, just because it requires a closed system, has 
an inherent conceptual problem when it comes to small systems of few particles.
Within the microcanonical ensemble theory one has 
\begin{equation}
T(E)=\left(\frac{\partial S(E)}{\partial E}\right)^{-1}_V.
\label{eq:muT}
\end{equation}
For macroscopic systems, it can be proven experimentally that $T$ from this 
equation equals the thermodynamical temperature defined by primary and 
secondary temperature standards, like the triple point of water. Unfortunately,
for small systems, this experimental proof can in principle not be performed 
due to the following difficulty. Temperature measurements involve necessarily 
energy exchange between the system and the thermometer in order to bring the 
thermometer to the same temperature as the system under study, thus undermining
the requirement of a closed system for a microcanonical description. For 
macroscopic systems, this problem is solved by introducing a small thermometer 
compared to the system. For small systems of a few particles, in general, every
thermometer will, due to its size, act as a heat bath when brought into contact
with the system under study. Thus, temperature measurements on small, truly
microcanonical systems seem virtually impossible and, therefore, it cannot be 
proven that the quantity $T$ from Eq.\ (\ref{eq:muT}) corresponds to the 
thermodynamical temperature in these cases, although $S$ and $E$ are still 
well defined and can be measured. It should also be noted that all means of 
temperature measurements which are based on the application of the statistical 
theory, like momentum distribution measurements, are not valid either, since 
the connection between the microcanonical statistical theory and thermodynamics
is always based on Eq.\ (\ref{eq:muT}).

The treatment of the nucleus in the canonical ensemble theory has also the
advantage that a classification scheme for phase transitions exists which 
reduces to the Ehrenfest definition in the thermodynamical limit 
\cite{BM00,MB01}. Hence, the difficulties mentioned in the Introduction can be 
overcome. Although it was suggested in \cite{Gr97,GV00} that the canonical 
ensemble theory should not be applied to first-order phase transitions, it has 
been shown that it gives convincing results in the case of nuclear 
multifragmentation \cite{MB01} which is thought to be a first-order phase 
transition. It has also been objected that the Laplace transformation involved 
in the canonical ensemble theory smears out the thermodynamical information of 
the system under study and that the canonical partition function therefore 
contains less information than the microcanonical partition function 
\cite{Gr97}. We want to emphasize, however, that the partition function of the 
canonical ensemble, when known on all of the right-hand half of the complex 
temperature plane, certainly does carry as much thermodynamical information of 
the system as the microcanonical partition function. The reason for this is 
that the inverse Laplace transformation (from the canonical to the 
microcanonical partition function) is mathematically defined in a unique 
fashion and hence the microcanonical partition function can in principle be 
recovered unanimously. However, since the inverse Laplace transformation 
requires integration on the complex temperature plane along a parallel to the 
imaginary axis, one has to expect that knowledge of the canonical partition 
function over all of the right hand half of the complex temperature plane is 
pivotal for revealing the complete thermodynamical information of the system 
under study. The classification scheme \cite{BM00,MB01} which we intend to 
apply in this work, is based on the DOZ of the canonical partition function and
therefore complies with this requirement.

\section{Model}

\begin{figure}\centering
\includegraphics[totalheight=17.9cm]{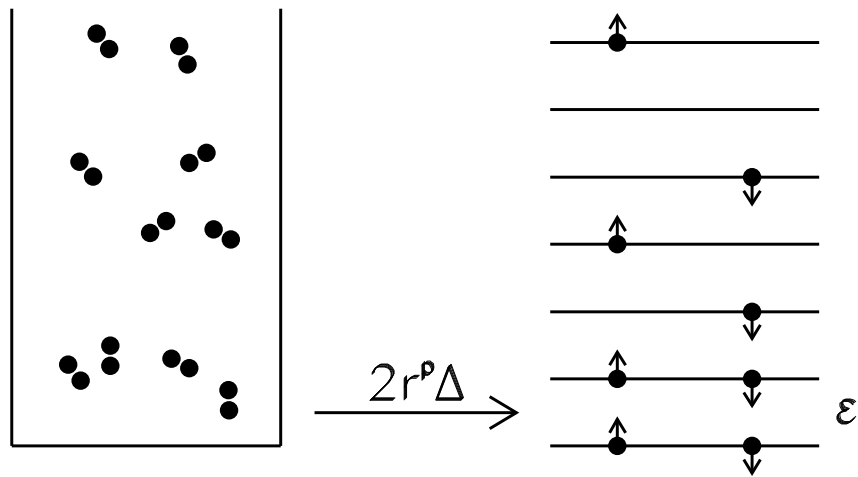}
\caption{Schematic representation of the nuclear model. For details, see text.}
\label{fig:model}
\end{figure}

The model in this work is a further development of the previously proposed
model in \cite{GH01a,GH01b}. We will first recapitulate those features of the 
old model which we have adopted in the present work and then describe the 
changes and additions. The basic idea behind our model is the assumption of a 
reservoir of nucleon pairs. These nucleon pairs can be broken and the unpaired 
nucleons are then promoted into an infinite, equidistant, doubly degenerated 
single-particle level scheme. The nucleon pairs in the reservoir do not 
interact with each other and are thought to occupy an infinitely degenerated 
ground state. The nucleons in the single particle level scheme do not interact 
with each other either, but they have to obey the Pauli principle. The model is
depicted in Fig.\ \ref{fig:model}. The essential parameters of the model are: 
the number of pairs in the reservoir at zero temperature $N$, the spacing of 
the single-particle level scheme $\epsilon$, and the energy necessary to break 
a nucleon pair $2\Delta$. Quenching of pairing correlations is introduced into 
this model by reducing the required energy to break a nucleon pair in the 
presence of unpaired nucleons. We assume that for every already broken nucleon 
pair, the energy to break a further nucleon pair is reduced by a factor 
$r\leq 1$. Thus, in the presence of $p$ broken nucleon pairs, the energy to 
break the $(p+1)$th nucleon pair is reduced to $2\,r^{p}\Delta$. We assume 
further that the single unpaired nucleon in an odd nucleus reduces the 
necessary energy for breaking the first nucleon pair of the same species to 
$2\,\sqrt{r}\Delta$.

The proposed model is supposed to display three distinct phases: a phase where
pairing correlations dominate, a phase where the pairs have essentially been
broken up and the system behaves like a non-interacting Fermi-gas dominated by 
Pauli-blocking, and a phase where the nucleons become diluted in a phase space
much larger in size (in terms of the number of realizations of thermal 
excitations) than the number of nucleons present. In this third phase, the 
Pauli blocking can be neglected and the system resembles mostly a classical 
ideal gas. We will, in the following, denote the three phases as the paired 
phase, the unpaired phase, and the quasi-classical phase. The three distinct 
phases of the system makes it possible to investigate two different phase 
transitions with our model.

It has been shown previously \cite{GH01a}, that the partition function for $n$ 
unpaired nucleons in an infinite, equidistant, doubly degenerated 
single-particle level scheme can be written as
\begin{eqnarray}
z_n&=&\sum_{i=0}^{(n-1)/2}2\,z_{n-i}^\uparrow\,z_i^\uparrow\hspace*{5mm}
n={\mathrm{odd}}\\
z_n&=&\sum_{i=0}^{(n-2)/2}2\,z_{n-i}^\uparrow\,z_i^\uparrow+
(z_{n/2}^\uparrow)^2\hspace*{5mm}n={\mathrm{even}},
\end{eqnarray}
where the $z_i^\uparrow$ are the partition functions for $i$ spin-up nucleons 
in an infinite, equidistant, non-degenerated single-particle level scheme. The 
$z_i^\uparrow$ can be expressed by means of the recursive formula \cite{GH01a}
\begin{eqnarray}
z_0^\uparrow&=&1\\
z_{i+1}^\uparrow&=&z_i^\uparrow\frac{\exp(-i\epsilon\beta)}
{1-\exp(-(i+1)\epsilon\beta)},
\end{eqnarray}
where $\beta=1/T$\footnote[1]{Here, and in the following, we set Boltzmann's 
constant $k_{\mathrm{B}}=1$.}. According to our model, in order to promote 
$n$ nucleons from the reservoir into the single-particle level scheme, one has 
to break $p=n/2$ nucleon pairs which requires the energy
\begin{equation}
2\Delta+2\,r\Delta+2\,r^2\Delta+\ldots+2\,r^{p-1}\Delta=2\Delta
\frac{1-r^p}{1-r}
\end{equation}
in the even case. In the odd case, one has to multiply this energy by 
$\sqrt{r}$ in order to take into account the quenching effect of the odd 
nucleon. Hence, the partition function for $n$ unpaired nucleons in our model 
becomes
\begin{eqnarray}
Z_n&=&z_n\,\exp(-2\,\sqrt{r}\Delta\frac{1-r^{(n-1)/2}}{1-r}\beta)\hspace*{5mm}
n={\mathrm{odd}}\\
Z_n&=&z_n\,\exp(-2\Delta\frac{1-r^{n/2}}{1-r}\beta)\hspace*{5mm}
n={\mathrm{even}}.
\end{eqnarray}
In order to obtain the partition function for the complete system, one has to 
sum over all possible numbers of unpaired nucleons in the single-particle level
scheme. Thus, the partition function of our model is finally
\begin{eqnarray}
Z_{\mathrm{pair}}&=&\sum_{n=1}^{N+1}Z_{2n-1}\hspace*{5mm}{\mathrm{odd}}\\
Z_{\mathrm{pair}}&=&\sum_{n=0}^NZ_{2n}\hspace*{5mm}{\mathrm{even}}.
\end{eqnarray}
At this point, the average number of unpaired particles can be calculated in a 
straightforward manner
\begin{eqnarray}
\langle n\rangle&=&\sum_{n=1}^{N+1}n\,Z_{2n-1}/\sum_{n=1}^{N+1}Z_{2n-1}
\hspace*{5mm}{\mathrm{odd}}\\
\langle n\rangle&=&\sum_{n=0}^Nn\,Z_{2n}/\sum_{n=0}^NZ_{2n}\hspace*{5mm}
{\mathrm{even}}.
\end{eqnarray}
The number of unpaired particles will fluctuate, and the fluctuations can be
calculated by
\begin{equation}
\delta n=\sqrt{\langle n^2\rangle-\langle n\rangle^2},
\end{equation}
where $\langle n^2\rangle$ is given by
\begin{eqnarray}
\langle n^2\rangle&=&\sum_{n=1}^{N+1}n^2\,Z_{2n-1}/\sum_{n=1}^{N+1}Z_{2n-1}
\hspace*{5mm}{\mathrm{odd}}\\
\langle n^2\rangle&=&\sum_{n=0}^Nn^2\,Z_{2n}/\sum_{n=0}^NZ_{2n}\hspace*{5mm}
{\mathrm{even}}.
\end{eqnarray}

Since there are protons and neutrons in the nucleus, one has to create two
partition functions $Z_{\mathrm{pair}}$, one for every nucleon species, and 
multiply them with each other. Here, one can introduce different parameters 
$\epsilon$, $\Delta$, $r$, and $N$ for protons and neutrons. For the sake of 
simplicity, we will, however, in this work not make use of this 
possibility\footnote[2]{This means that the even-odd and the odd-even systems 
will be equal. We will, therefore, in this work use the term 'odd system' to 
denote either of the two.}. In addition to the model partition function, we 
have in the previous work \cite{GH01b} also assumed a rotational and 
vibrational partition function
\begin{eqnarray}
Z_{\mathrm{rot}}&=&\sum_{I=0,2,4\ldots}^{12}
\exp(-A_{\mathrm{rig}}I(I+1)\beta)\\
Z_{\mathrm{vib}}&=&\sum_{\nu=0}^13^\nu\exp(-\nu\hbar\omega_{\mathrm{vib}}\beta)
\end{eqnarray}
with $A_{\mathrm{rig}}$ being the rigid-body rotational parameter and 
$\omega_{\mathrm{vib}}$ the vibrational frequency of the nucleus under study.
The nuclear partition function becomes therefore in the most general case
\begin{equation}
Z=Z_{\mathrm{pair}}^\pi\,Z_{\mathrm{pair}}^\nu\,Z_{\mathrm{rot}}\,
Z_{\mathrm{vib}}
\end{equation}
where we will use in this work equal proton and neutron partition functions 
$Z_{\mathrm{pair}}^\pi$ and $Z_{\mathrm{pair}}^\nu$. We would like to stress 
that for $r=1$, the present model is approximately equal to the previous model 
\cite{GH01a,GH01b}, whereas the possibility of $r<1$, i.e., the quenching of 
pairing correlations, is a new feature of the present model.

Although the model is purely phenomenological, we would like to emphasize that
it can be closely related to a microscopical model. If one assumes a 
Hamiltonian of the form
\begin{equation}
\hat{H}=\hat{H}_\pi+\hat{H}_\nu+\hat{H}_{\mathrm{rot}}+\hat{H}_{\mathrm{vib}}
\end{equation}
with 
\begin{equation}
\hat{H}_{\pi,\nu}=\epsilon\sum_\kappa a^\dagger_\kappa a_\kappa
-|G|\sum_{\kappa,\lambda>0}a^\dagger_\kappa a^\dagger_{-\kappa}a_{-\lambda}
a_\lambda,
\end{equation}
i.e., a simple single-particle plus pairing Hamiltonian for protons and 
neutrons, and conventional rotational and vibrational Hamiltonians 
$\hat{H}_{\mathrm{rot}}$ and $\hat{H}_{\mathrm{vib}}$ for the collective modes,
similar results as in this work can be achieved \cite{BD01,GB00}. The advantage
of the present, phenomenological model over the microscopic model is that, in
general, a much larger number of particles and single-particle levels can be 
taken into account.

\section{Model properties}

It has already been shown that the previous model \cite{GH01a,GH01b} describes 
well the level density and heat capacity in the $^{162}$Dy nucleus. Further, it
could reproduce \cite{GH01b} the anchor points of the level density curve 
proposed in \cite{GH01a} for several other mid-shell nuclei. However, 
in order to do so, one had to assume values for $\epsilon$ which were off 
general estimates of the shell model. Also, it was not possible to reproduce 
the theoretical heat capacity curve of iron nuclei as calculated in \cite{LA01}
within the SMMC approach. Since in the new model an additional parameter is 
introduced, we have the hope to use more reasonable values of $\epsilon$ and 
still be able to describe the experimental data with good accuracy. 

\begin{figure}\centering
\includegraphics[totalheight=17.9cm]{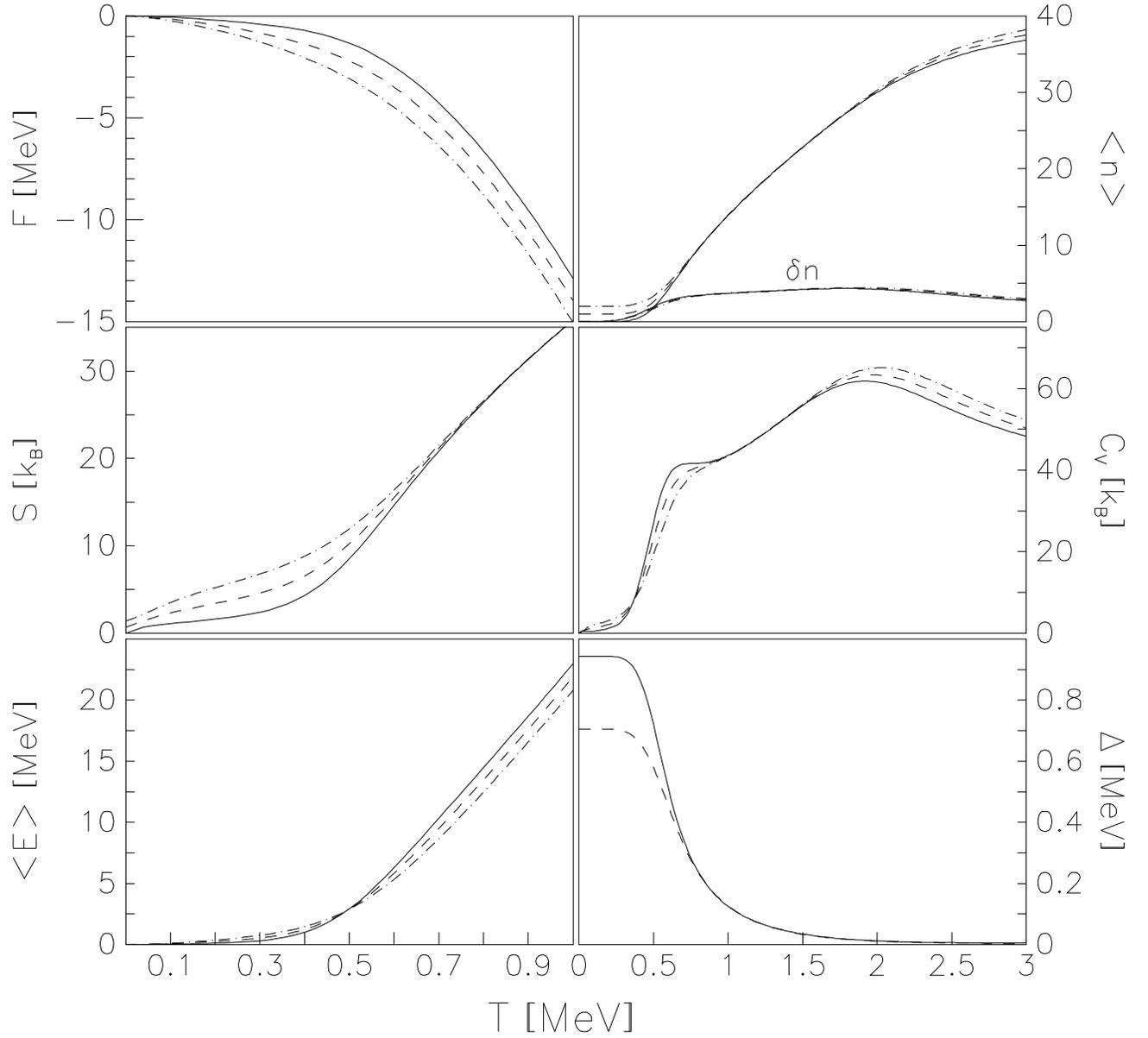}
\caption{Thermodynamical quantities calculated within the nuclear model. The 
solid, dashed, and dash-dotted lines are for the even-even, odd, and odd-odd 
system, respectively. For $\Delta (T)$, only the even and odd case for one 
nucleon species is shown. The long tail of the $\Delta(T)$ curves at high 
temperatures agrees well with modern RPA calculations [38].
The parameters are those for $^{162}$Dy.}
\label{fig:thermo}
\end{figure}

The parameters for the present nuclear model are taken from the following
systematics \cite{BM69}
\begin{equation}
\epsilon=\frac{3\pi^2}{A}\,{\mathrm{MeV}},
\label{eq:epsilon}
\end{equation}
where the level density parameter $a=\pi^2/3\epsilon$ is related to the mass
number $A$ by $a=A/9\,{\mathrm{MeV}}^{-1}$. The pairing-gap parameter $\Delta$ 
can also be related to the mass number $A$ by
\begin{equation}
\Delta=\frac{12}{\sqrt{A}}\,{\mathrm{MeV}}.
\end{equation}
The rigid-body rotational parameter $A_{\mathrm{rig}}$ can be expressed in 
terms of the rigid moment of inertia $\Theta_{\mathrm{rig}}$ by 
\begin{equation}
A_{\mathrm{rig}}=\hbar^2/2\,\Theta_{\mathrm{rig}}.
\end{equation}
Assuming the nucleus to be a rotational symmetric ellipsoid of constant 
density, the rigid moment of inertia can be expressed by the nuclear mass $M$ 
and mean radius $R$ as $\Theta_{\mathrm{rig}}=2/5\,MR^2$. The nuclear mean 
radius can then again be related to the mass number by 
$R=1.24\,{\mathrm{fm}}\,A^{1/3}$. The vibrational frequency of the nucleus can 
be taken directly from spectroscopic data \cite{FS96}. For the parameter $r$ of
our model, which governs the evolution of the pairing gap $\Delta$ with 
temperature, no systematics exist. Theoretical studies suggest a quenching of 
the pairing energy per broken pair in the order of 0.56 in the mass 190 region 
\cite{DK95b} yielding a $\Delta(T)$ curve comparable to results from modern RPA
calculations \cite{DZ01} (see Fig.~\ref{fig:thermo}). Probably $r$ will 
increase with mass number. This effect is, however, not taken into account in 
this work and we adopt $r=0.56$ for all nuclei studied. Finally, for the last 
parameter $N$ of the model, we assume ten neutron and ten proton pairs in the 
reservoir at zero temperature in most of this work. This is an increase over 
the previous calculations \cite{GH01a,GH01b} where a total number of ten 
nucleon pairs was assumed. The increase of the number of nucleon pairs in the 
reservoir allows us to map thermodynamical quantities of nuclei up to higher 
temperatures of, e.g., 1.5~MeV in dysprosium. A summary of the applied 
parameters is listed in Table~\ref{tab:parameters}.

In order to compare the model to experimental data, one has to calculate 
experimentally observed quantities like the free energy $F$, the entropy $S$, 
the caloric curve $\langle E(T)\rangle$, the heat capacity $C_V$, and the 
nuclear level density $\rho(\langle E\rangle)$ from the canonical partition 
function. While the first four quantities can be calculated in a 
straightforward fashion by the following expressions
\begin{eqnarray}
F&=&-T\ln Z\\
S&=&-\left(\frac{\partial F}{\partial T}\right)_V\\
\langle E\rangle&=&F+ST\\
C_V&=&\left(\frac{\partial\langle E\rangle}{\partial T}\right)_V,
\end{eqnarray}
the nuclear level density should, in principle, be obtained by an inverse 
Laplace transformation of the canonical partition function. We will, however, 
in this work make use of the saddle-point approximation \cite{NA97}
\begin{equation}
\rho(\langle E\rangle)=\frac{\exp (S)}{T\sqrt{2\pi C_V}},
\end{equation}
which gives satisfactory results for the nuclear level density \cite{GH01b}.
Finally, we will calculate the average number of unpaired nucleons 
$\langle n\rangle$, the fluctuation of this number $\delta n$, and the 
effective, temperature-dependent pairing energy 
$\Delta (T)=r^{\langle n/2\rangle}\Delta (0)$.

In Fig.~\ref{fig:thermo}, thermodynamical quantities derived from the model are
shown using parameters for $^{162}$Dy. The quantities $S$, $\langle n\rangle$, 
$C_V$ and $\Delta (T)$ are coinciding for the even-even, odd, and odd-odd 
systems above temperatures of 0.8--1~MeV, indicating that the pairing 
correlations are already completely quenched at these temperatures. Indeed, the
heat capacity curve displays a linear dependence on temperature between $T=1$ 
and 1.5~MeV, which is the expected temperature dependence for a Fermi gas. The 
fluctuation of the number of unpaired particles $\delta n$ increases strongly 
around $T\sim 0.5$~MeV, which coincides with a strong decrease of the effective
pairing energy $\Delta (T)$ and the S-shape of the heat capacity curve, first 
observed experimentally in \cite{SB01}. The latter three observations can be 
interpreted as signatures of a second-order phase-transition like phenomenon in
nuclei. 

On the right-hand panels of Fig.~\ref{fig:thermo}, the quantities 
$\langle n\rangle$ and $C_V$ can be seen to split up again for the even-even, 
odd, and odd-odd systems at temperatures above 1.5--2~MeV\@. At these 
temperatures, all nucleons start to become excited, the reservoir is rapidly 
being exhausted and the system can no longer absorb energy at the same rate 
without getting heated up rapidly. Therefore, the heat capacity decreases until
it reaches the limit of $1\,k_{\mathrm{B}}$ per particle\footnote[3]{This 
corresponds to the value for a two-dimensional ideal, classical gas.} and the 
number of unpaired particles will increase at a progressively smaller pace to 
the limit of 40, 41, and 42 for the even-even, odd, and odd-odd systems, 
respectively. This is the signal of a possible second phase transition inherent
in the model from an unpaired phase to a quasi-classical phase. It is not 
expected that this second phase transition is present in the experimental data 
since the nuclei under study in this work have more particles than are included
in the model space. Therefore, the second phase transition in the model merely 
reflects the fact that the model space is still too small in order to describe 
heavy nuclei adequately at the highest temperatures\footnote[4]{In any case, 
atomic nuclei become unbound, undergo multifragmentation and therefore loose 
their identities at the critical temperature for the liquid-gas phase 
transition. This will therefore be the temperature beyond which our model 
becomes unrealistic, even when including all nucleons of the nucleus.}. Some 
kind of signal of the exhaustion of the model space will, however, be present 
in all types of nuclear model calculations which do not include all nucleons 
or the complete single-particle level scheme. We will therefore investigate in
Sect.~\ref{sec:results} whether and how the truncation of the model space 
interferes with the signal of the pairing phase transition. 

\begin{figure}\centering
\includegraphics[totalheight=17.9cm]{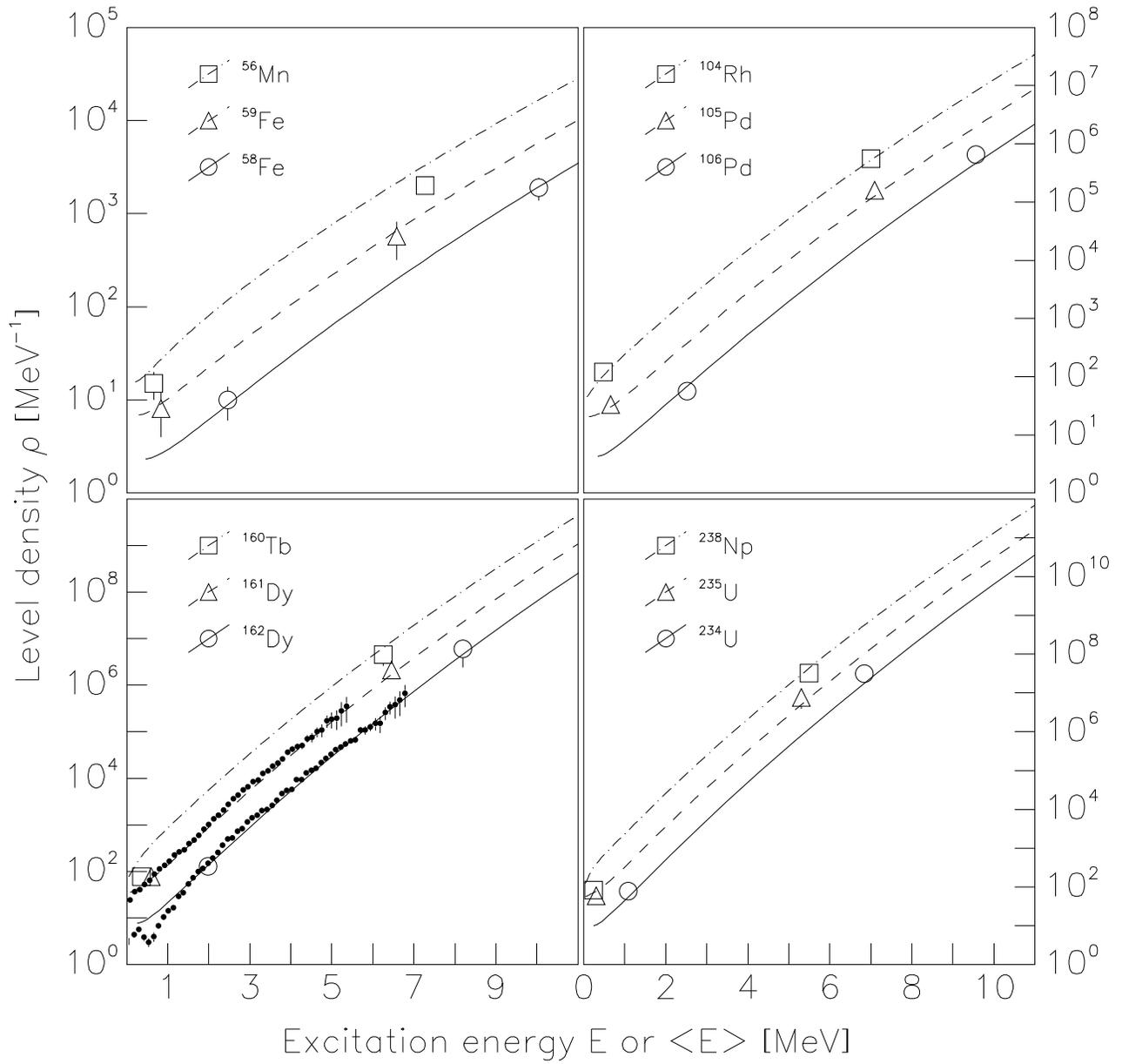}
\caption{Comparison of calculated and experimental level density. The open 
symbols are the anchor points of Ref.~[32],
the full symbols are the experimental level densities of Ref.~[27]
using a slightly different normalization (see text).}
\label{fig:levdens}
\end{figure}

Figure~\ref{fig:levdens} shows how the model compares to the level-density
anchor points of Ref.~\cite{GH01a} and the measured level densities of two
dysprosium isotopes \cite{SB01}. The data points for the dysprosium isotopes
have been modified slightly for the purpose of this work. While in the previous
works \cite{SB01,SB00} a back-shifted level-density formula with the von~Egidy
parameterization \cite{ES87} was used for normalization and extrapolation to 
the neutron-binding energy and beyond, in this work, we will make use of a 
simple, shifted level-density formula 
\begin{equation}
\rho=f\frac{\exp (2\sqrt{aU})}{12\sqrt{0.1776}\,a^{1/2}U^{3/2}A^{1/3}}
\end{equation}
with $U=E-\Delta$, $a=A/9$~MeV$^{-1}$ and $\Delta=12/\sqrt{A}$~MeV for 
consistency. The parameterization for the spin cut-off parameter $\sigma$ is
taken from Ref.~\cite{GC65} and the normalization parameter $f$ is chosen such
that the level-density formula intersects the data point from neutron resonance
spacing \cite{IA98}. It is satisfying that the model reproduces the data points
well without any fine tuning of the parameters in the different mass regions
off closed shells. Only in the case of the iron mass region, the single 
particle level spacing had to be increased from the systematic value of
$\epsilon=517$~keV to 800~keV in order to describe the data. The shortcomings 
of the model to reproduce the iron data using parameters from the systematics 
is most likely due to the presence of the double shell closure in the 
neighboring $^{56}$Ni isotope. Irregularities of both, the single-particle 
level spacing (or more precisely the closely related level-density parameter 
$a$) and the pairing-gap parameter $\Delta$ in the vicinity of closed shells 
are a well-known fact \cite{BM69}. We would like to stress that a better fit to
the experimental data might certainly be achieved by fine tuning of the 
parameters for protons and neutrons for every mass region. However, the 
possibility of using systematic parameters for the model highlights its general
applicability.

\begin{figure}\centering
\includegraphics[totalheight=17.9cm]{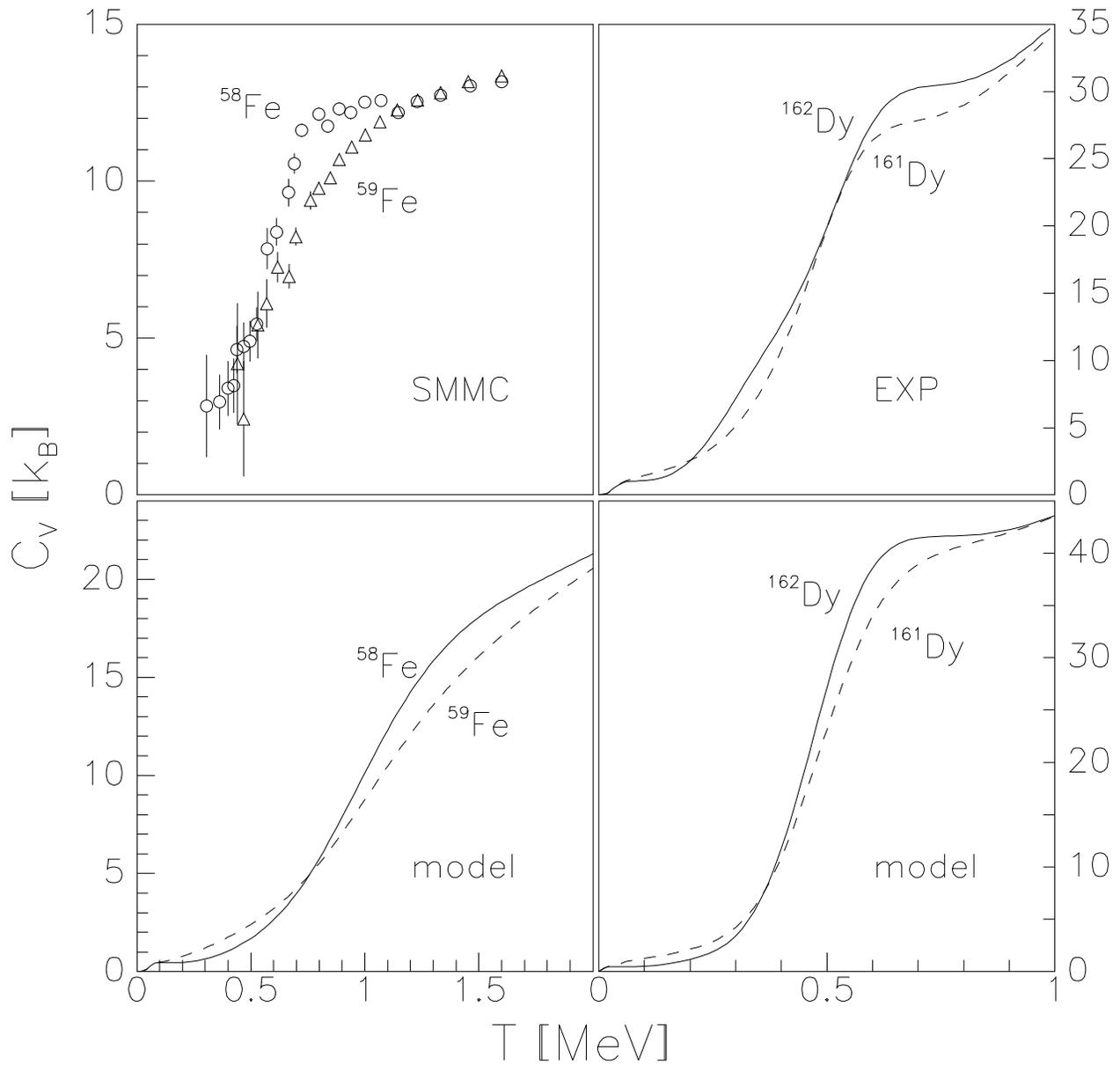}
\caption{Comparison of heat capacity curves of the present model and of 
other publications. The open symbols are scanned from Ref.~[26],
the data for the dysprosium nuclei are from Ref.\ [27]
using a slightly different normalization (see text).}
\label{fig:cv}
\end{figure}

In Fig.~\ref{fig:cv}, the model predictions are compared to published canonical
thermodynamical quantities. For the iron nuclei these data are recent SMMC 
calculations \cite{LA01}, while for the dysprosium nuclei, semi-experimental
data exist \cite{SB01}. The comparison is made for canonical heat-capacity
curves, since these curves are effectively second derivatives of the canonical
partition functions and therefore more sensitive to structural changes in the 
nucleus than any other thermodynamical quantities. 

Although the steepness of the S-shape differs somewhat and the absolute value 
at the plateau is off by $\sim 30$\%, the overall agreement of the model 
calculation for the dysprosium nuclei with the semi-experimental data is 
satisfying. Especially the temperature where the steepness of the S-curve 
reaches the maximum, i.e., the critical temperature of the pairing phase 
transition, is well reproduced. For the iron nuclei, the agreement between our 
model and the SMMC calculations is not so good. Unfortunately, no experimental 
data exists to compare the two models to, therefore it is difficult to judge 
which is the more realistic one. While the present model is very schematic, it 
offers the possibility to include infinitely many single-particle levels and 
much more nucleons than the SMMC calculations. Especially the necessity of a 
finite single-particle level scheme in SMMC calculations will cause the heat 
capacity curve to approach zero for high temperatures and it is not clear at 
which temperature this effect takes over in the data of Ref.~\cite{LA01}. On 
the other hand, the SMMC calculation can incorporate shell-effects in a more 
natural way than by simply adjusting the two parameters $\epsilon$ and 
$\Delta$. Especially the use of a realistic nucleon-nucleon interaction and the
correct single particle energies for the complete $(pf+0g_{9/2})$ shell might 
change the picture significantly at low temperatures compared to the present 
schematic model.

In conclusion, it can be stated that the present model, using a systematic 
parameter set, reproduces fairly well the available experimental data, but a 
discrepancy with SMMC calculations in the iron-mass region exists. Since the 
agreement with experiment is good in the rare earth mass region, where also 
experimental data is abundant, we are confident that the model can be applied 
to classify the order of the pairing phase transition with good accuracy in 
deformed rare earth nuclei if not in most mid-shell nuclei.

\section{Classification of phase transitions}

\subsection{Classification scheme}

The classification scheme for phase transitions is based entirely on 
\cite{BM00,MB01} and is only slightly modified for the purpose of this work. 
We will therefore only recapitulate briefly the main features of the 
classification scheme. The scheme relies on the DOZ and derives three 
quantities, where two of them classify the order of a possible phase transition
and the third one reflects the discreteness of the system under study. 

\begin{figure}\centering
\includegraphics[totalheight=17.9cm]{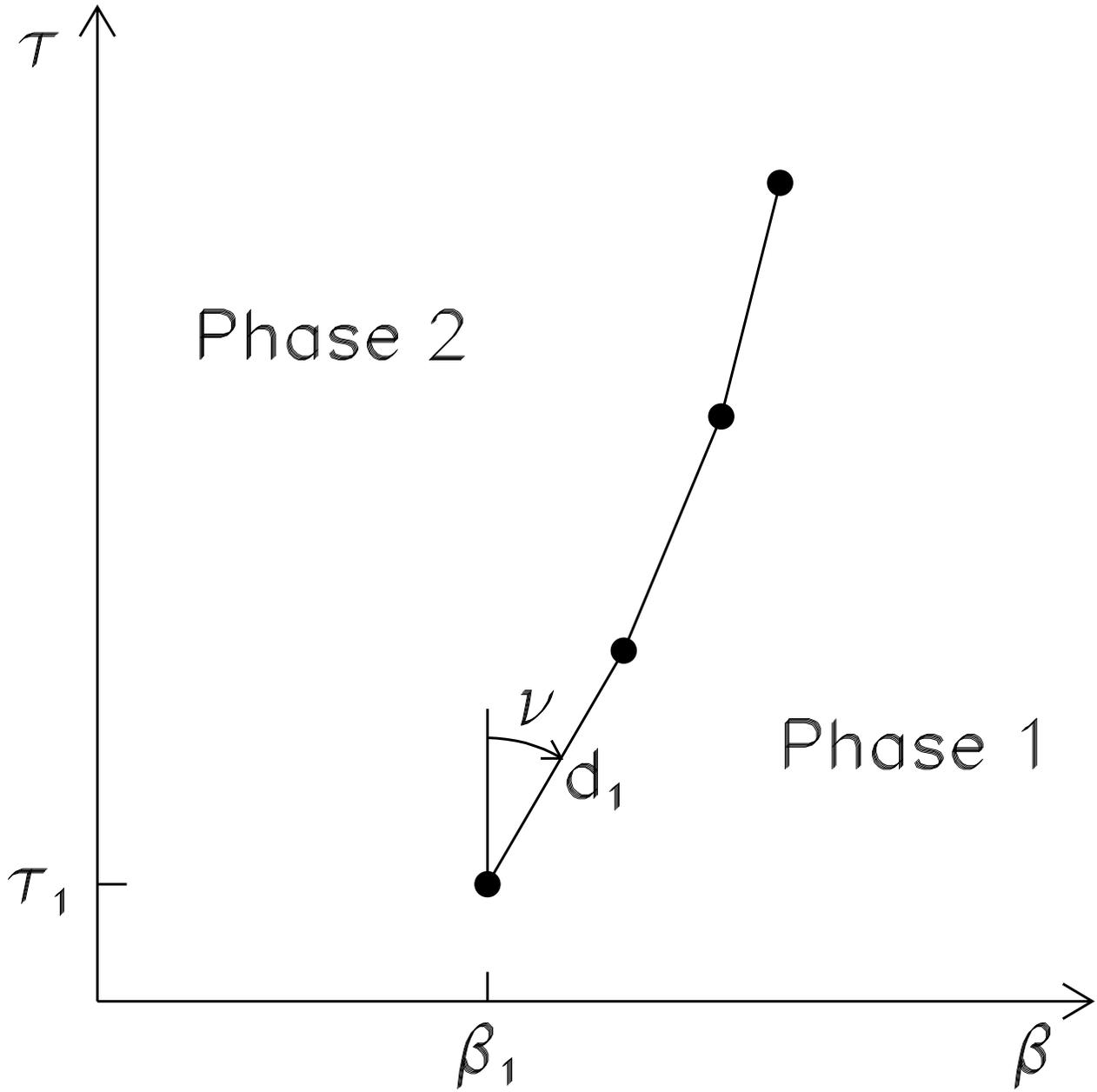}
\caption{Definition of important quantities for the classification of phase
transitions in finite systems.}
\label{fig:method}
\end{figure}

First, we define the inverse complex temperature
\begin{equation}
{\cal{B}}=\beta+i\tau,
\end{equation}
where $\beta=1/T$ as usual and $\tau$ denotes the imaginary part of the inverse
complex temperature which is measured in MeV$^{-1}$ in this work. The zeros of 
the canonical partition function typically line up on curves through the 
complex temperature plane. Whenever such a curve intersects the real axis, a 
phase transition in the system might occur. Figure~\ref{fig:method} shows a 
typical example with four zeros on a curve approaching the real axis. The zeros
are denoted by $(\beta_j,\tau_j)$ with $j=1\ldots 4$ and $j$ increasing with 
increasing distance from the real axis. 

The angle $\nu$, being the first quantity of interest, is calculated by
\begin{equation}
\nu=\arctan\frac{\beta_2-\beta_1}{\tau_2-\tau_1}.
\end{equation}
Further, the average inverse distance between zeros can be calculated as 
\begin{equation}
\Phi=\frac{1}{2}\,\left(\frac{1}{d_{j-1}}+\frac{1}{d_j}\right),
\end{equation}
where $d_{j}=\sqrt{(\beta_{j+1}-\beta_j)^2+(\tau_{j+1}-\tau_j)^2}$. The 
function $\Phi$ can then be approximated in the vicinity of the real axis by a 
power law of $\tau_j$ only
\begin{equation}
\Phi(\tau_j)\propto\tau_j^\alpha,
\end{equation}
such that the second quantity of interest, $\alpha$, can be calculated by means
of
\begin{equation}
\alpha=\frac{\ln\Phi(\tau_3)-\ln\Phi(\tau_2)}{\ln\tau_3-\ln\tau_2}.
\end{equation}
One can easily see that in order to calculate $\alpha$, one has to determine at
least the first four zeros closest to the real axis. With a slight redefinition
of the function $\Phi$ by
\begin{equation}
\tilde{\Phi}(\tilde{\tau}_j)=\frac{1}{d_j},
\end{equation}
where $\tilde{\tau}_j=(\tau_j+\tau_{j+1})/2$, only the first three zeros have 
to be known. The order of the phase transition is now completely determined by 
the two quantities $\nu$ and $\alpha$. When $\alpha<0$ or $\alpha=\nu=0$ the 
system exhibits a first-order phase transition. For $0<\alpha<1$ and $\nu=0$ or
$\nu\neq 0$, a second-order phase transition takes place. For $\alpha>1$ the 
phase transition is of higher order. 

The third parameter $\tau_1$ indicates the discreteness of the system. More
precisely, the quantity $\tau_1/\hbar$ equals the time after which an ensemble
of equal systems loses its memory. Only if $\tau_1\rightarrow 0$ for increasing
particle numbers, one has a corresponding phase transition in the Ehrenfest 
sense for the thermodynamical limit. For finite systems, one can then speak of 
a phase transition in the generalized Ehrenfest sense. Finally, $\beta_1$ might
be defined as the inverse of the critical temperature of the phase transition. 
In the following, the classification scheme will be applied to the DOZ of 
mid-shell nuclei calculated within the proposed nuclear model.

\subsection{Results}
\label{sec:results}

\begin{figure}\centering
\includegraphics[totalheight=22.3cm]{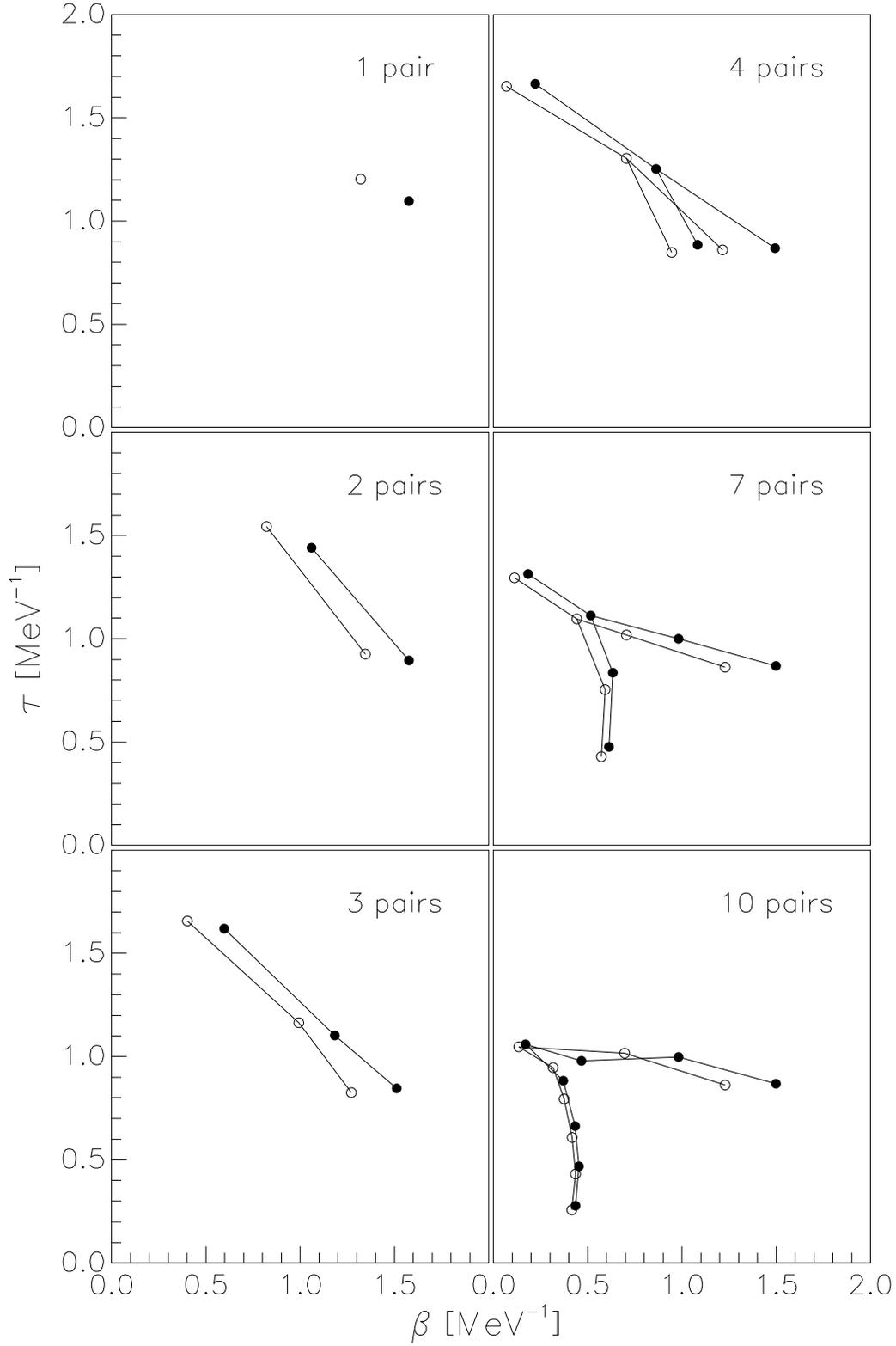}
\caption{DOZ for different numbers of pairs in the reservoir. The full symbols 
are for the even system, the open symbols for the odd system. The parameters of
the model are those for $^{162}$Dy. There exist zeros with $\tau>2$~MeV$^{-1}$ 
and/or $\beta<0$~MeV$^{-1}$, but these are not of interest in this context.}
\label{fig:pairs}
\end{figure}

First, we investigate the effect of the numbers of pairs in the reservoir at 
zero temperature on the DOZ. On the upper left panel in Fig.~\ref{fig:pairs},
the calculation is done for one single pair in the reservoir. Certainly, no 
phase transition can be claimed, but the positions of the zeros already 
indicate at which temperatures the pairing phase transition will take place. 
Also, the effect of the Pauli blocking by the single unpaired nucleon is 
visible, shifting the zero of the odd system to higher temperatures. For two 
pairs in the reservoir, two zeros for the even and odd system can be seen. The 
zeros are positioned on slanted lines where the zeros of the odd system are 
again shifted to higher temperatures due to the Pauli blocking. The same is 
true for three pairs in the reservoir. According to the previous subsection, 
three zeros in a line can already be used to classify a phase transition. In 
the present case, since $\alpha$ is negative, a first-order phase transition is
indicated. Due to the small size of the system, however, one cannot attribute 
this phase transition simply to the quenching of pairing correlations. Rather, 
one has to acknowledge the fact that the pairing phase transition coincides 
with the phase transition from an unpaired to a quasi-classical system, giving 
rise to a disturbed signal for the order of the phase transition. This becomes 
clearer in the case of four pairs in the reservoir. Here, one might already 
discuss two separate phase transitions, since one can distinguish two different
curves of zeros approaching the real axis. While the curves to the right might 
be attributed to the pairing phase transition, the curves to the left are 
indicating the transition to a quasi-classical phase. Still, since the two 
curves are sharing zeros one cannot expect to get a clean signal for the 
classification of the two phase transitions and indeed for both curves one 
obtains negative values of $\alpha$, indicating two first order phase 
transitions, a result which is at variance with expectations. First the 
calculation with seven pairs separates the two phase transitions enough in 
temperature to get clean signals. Here, one finds negative values of $\alpha$ 
for the transitions to a quasi-classical phase, indicating a first-order phase 
transition. Interestingly, this phase transition does show only a very small 
odd-even effect, which can in total be attributed to the by one higher particle
number in the odd case. The pairing phase transition, however, has a large 
odd-even dependence and yields $\alpha\sim 0.86$ for the even case, indicating 
a second-order phase transition and a value greater than one for the odd case, 
indicating a higher order phase transition. This result remains stable for 
larger particle numbers. The values of $\nu$ are consistent with the 
conclusions drawn above.

\begin{figure}\centering
\includegraphics[totalheight=17.9cm]{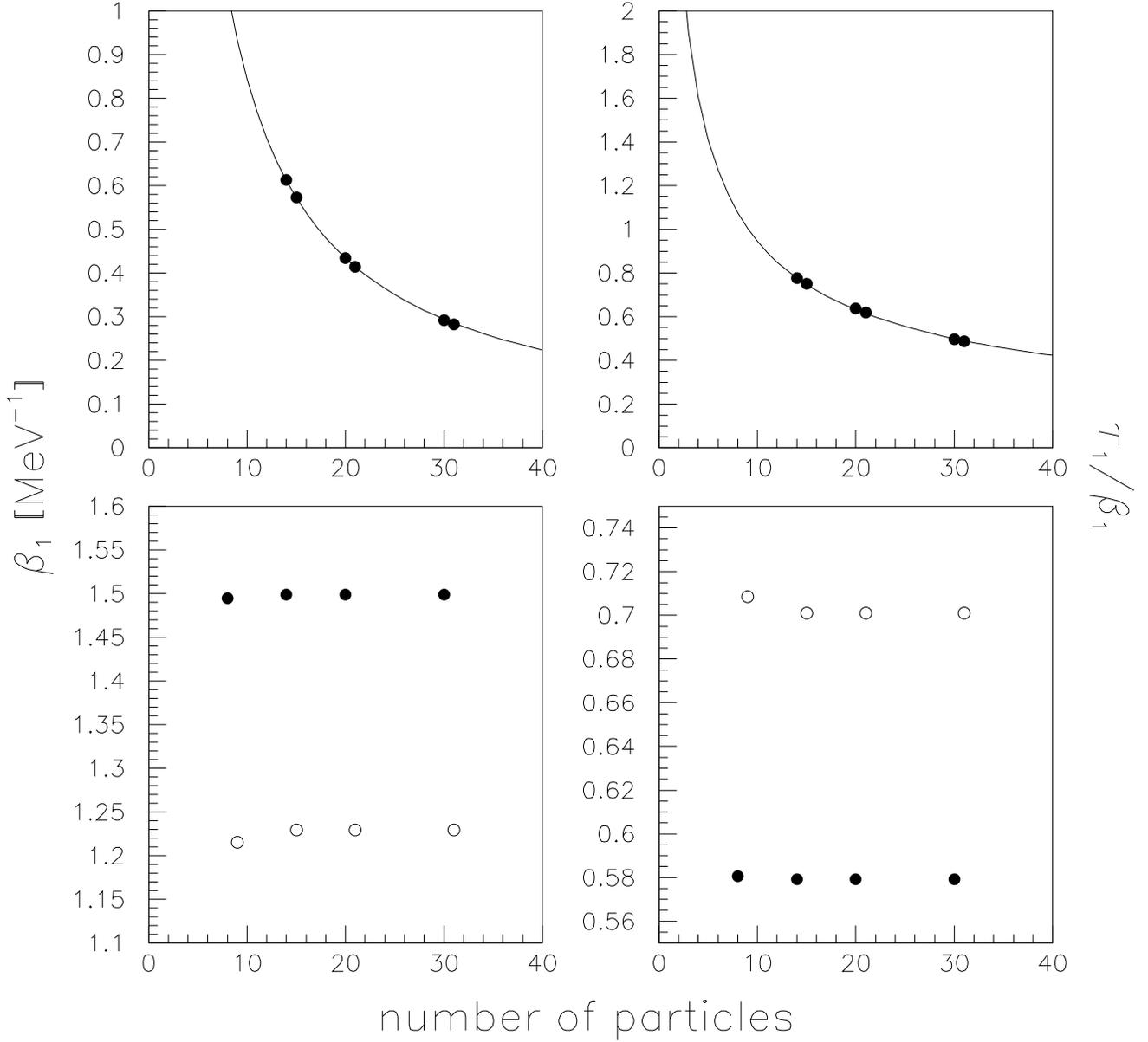}
\caption{Evolution of the parameters $\beta_1$ and $\tau_1/\beta_1$ as function
of the size of the model space for the two phase transitions. Upper panels: the
phase transition from the unpaired to the quasi-classical phase. Lower panels: 
the pairing phase transition. The full symbols in the lower panels are for the 
even system, the open symbols are for the odd system. The curves on the upper 
panels are fits to the data by a simple power law. The odd and even systems 
align on identical curves.}
\label{fig:beta-tau1}
\end{figure}

We will now investigate the evolution of the parameters $\beta_1$, i.e., the 
inverse critical temperatures of the respective phase transitions, as function 
of $N$, i.e., the size of the model space. In the upper left panel of 
Fig.~\ref{fig:beta-tau1}, it is shown that $\beta_1$ for the phase transition 
from the unpaired to the quasi-classical phase obeys a power law of the form 
$aN^b$ with $b\sim -1$. This means that in the thermodynamical limit the 
critical temperature of this phase transition will approach infinity and thus, 
the phase transition cannot be observed in infinitely large systems. This is 
not surprising, since the very nature of this phase transition, which is a 
transition from a Fermi gas to a quasi-classical phase where the nucleons are 
diluted in phase space, obviously requires increasingly higher temperatures the
more nucleons are present. Here, we want to stress once more that this phase 
transition is a true phase transition inherent in the model and might even be 
realized in nature by some physical system. However, when applying the model to
heavy nuclei which are much larger than the model space, $\beta_1$ merely 
indicates at which (inverse) temperature the reservoir is being exhausted and 
that the model space is too small to describe heavy nuclei adequately below 
$\beta_1$. Eventually, the temperature beyond which our model becomes 
unrealistic, even when including all nucleons of the nucleus in the model 
space, is given by the critical temperature for the liquid-gas phase transition
(see also \footnotemark[4]). 

On the lower left panel of Fig.~\ref{fig:beta-tau1}, the evolution of $\beta_1$
for the pairing phase transition as function of the size of the model space is 
shown. The inverse critical temperature $\beta_1$ of the pairing phase 
transition is independent of the size of the model space as it is expected, but
it shows a pronounced odd-even effect due to the Pauli blocking of the single 
unpaired nucleon. 

Finally, we will investigate the parameters $\tau_1$ as function of $N$ in 
order to gain information on whether the phase transitions observed in the 
finite system under study correspond to true phase transitions in the 
thermodynamical limit. Unfortunately, in the case of the first-order phase 
transition from the unpaired phase to the quasi-classical phase, $\beta_1$ and 
therefore also $\tau_1$ show a trivial scaling with $N$ which is clouding the
relevant, additional $N$ dependence of $\tau_1$. Therefore, one has to 
investigate the dimensionless parameter $\tau_1/\beta_1$ for a residual, 
non-trivial $N$ dependence in order to be able to conclude on the reality of 
the phase transition in the generalized Ehrenfest sense for the finite system. 
On the upper right panel of Fig.~\ref{fig:beta-tau1} the dependence of 
$\tau_1/\beta_1$ on $N$ for the phase transition from the unpaired to the 
quasi-classical phase is shown to obey the power law $aN^b$ with $b\sim -0.5$. 
Thus, the dimensionless parameter $\tau_1/\beta_1$ is shown to approach the 
real axis in the thermodynamical limit and the modeled phase transition 
satisfies the generalized Ehrenfest definition of a phase transition. 

For the pairing phase transition, one could in principle investigate $\tau_1$
directly as function of $N$, since $\beta_1$ is independent of $N$ and thus, no
trivial scaling of $\tau_1$ with $N$ will alter the results. However, to treat
everything on an equal footing, also in this case, we have investigated the 
dimensionless parameter $\tau_1/\beta_1$. Since $\tau_1/\beta_1$ shows no sign
of approaching the real axis (see the lower right panel of 
Fig.~\ref{fig:beta-tau1}), one has to ask the question whether the modeled 
pairing phase transition corresponds to any true phase transition in the 
thermodynamical limit.

In order to answer this question, we introduce the notion that, in the case of 
the pairing phase transition, the size of the system is not determined by the 
size of the reservoir $N$, but rather by the mass number $A$ and subsequently 
by the mass-number-dependent parameters $\epsilon$ and $\Delta$. Specifically, 
the parameter $g=\Delta/\epsilon$ scales as $\sqrt{A}$ and should thus be 
increased in order to model a larger system. This is in sharp contrast to the 
analysis in \cite{BD01}, where the authors assume that the simple increase of 
the number of particles $N$ in the calculation should model heavier nuclei.

\begin{figure}\centering
\includegraphics[totalheight=17.9cm]{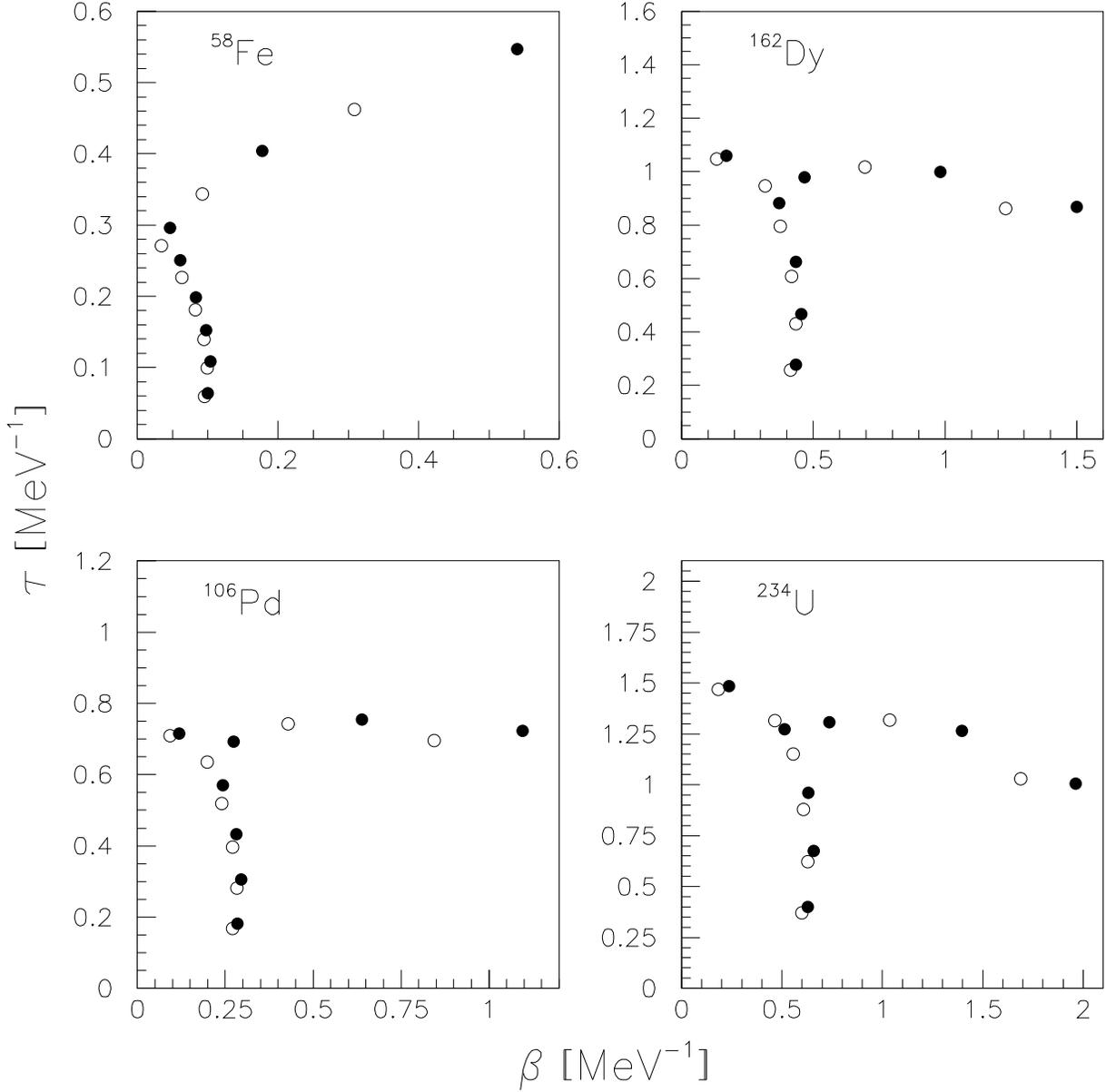}
\caption{DOZ for nuclei in four different mass regions. The full symbols are 
for the even systems, the open symbols stand for the odd systems. The line of 
zeros approaching the real axis perpendicularly corresponds to the exhaustion
of the finite model space and is not relevant in the discussion. Only the more 
or less horizontal lines of zeros reflect the quenching of pairing 
correlations.}
\label{fig:iso}
\end{figure}

\begin{figure}\centering
\includegraphics[totalheight=8.9cm]{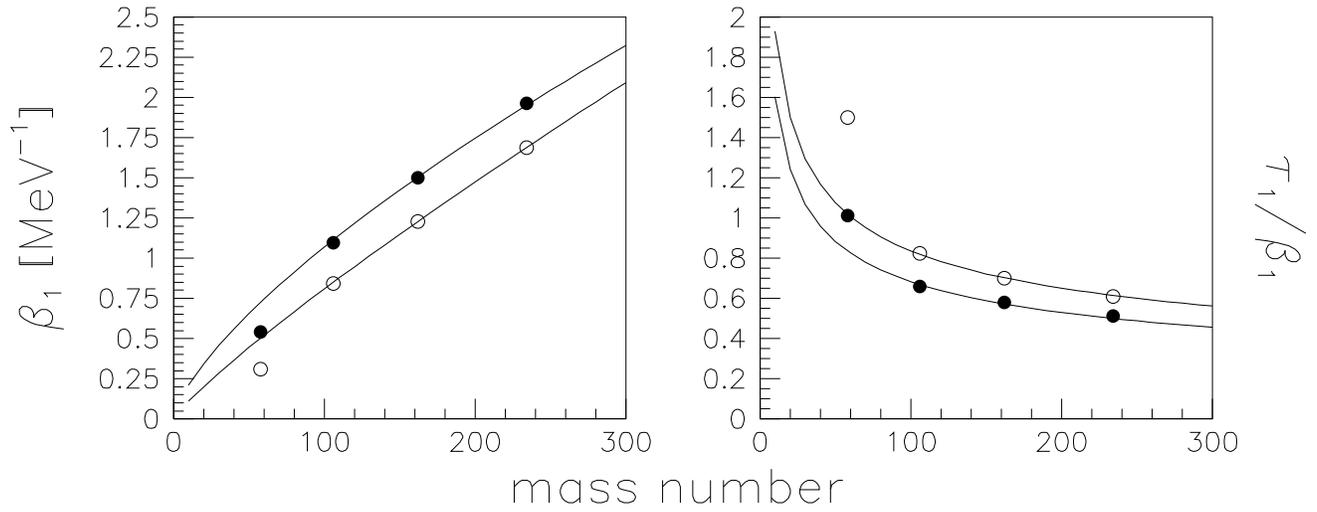}
\caption{Evolution of the parameters $\beta_1$ and $\tau_1/\beta_1$ of the
pairing phase transition with respect to the mass number $A$, i.e., the 
relevant size of the system under study. The full symbols are for the even 
systems, the open symbols stand for the odd systems. The points for 
$^{58,59}$Fe are not taken into account in the fit, since for the iron data, a
single-particle level spacing $\epsilon$ off the global systematics is used.}
\label{fig:beta-tau2}
\end{figure}

Therefore, in order to investigate the evolution of $\tau_1$ for the pairing 
phase transition with respect to the relevant size of the system, i.e., the 
mass number $A$, we have calculated the DOZ for nuclei in four different mass 
regions (see Fig.~\ref{fig:iso}). Obviously, the pairing phase transition can 
only be observed for nuclei with mass numbers $A>100$ within the presented 
model, since for lighter nuclei, the curve of zeros relevant for the quenching 
of pairing correlations points away from the real axis, and therefore, one 
cannot claim, within the model, the existence of a pairing phase transition for
light nuclei. On the left-hand panel of Fig.~\ref{fig:beta-tau2}, the inverse 
critical temperature of the pairing phase transition is shown to scale 
approximately with $A^{2/3}$ in the rare-earth mass region, which is expected 
when assuming $T_c=\Delta/S_1$, where $S_1$ is the single-particle entropy 
\cite{GH01a}. The iron data do not quite follow this trend, since we have used 
a single-particle level spacing for the iron nuclei outside the global 
systematics of Eq.~(\ref{eq:epsilon}), see also Table \ref{tab:parameters}. 
Again, by investigating the dimensionless parameter $\tau_1/\beta_1$ of the 
pairing phase transition, the trivial mass dependence is taken out and the 
residual dependence of $\tau_1/\beta_1$ on $A$ will tell if the modeled pairing
phase transition in the finite system under study is a phase transition in the 
generalized Ehrenfest sense. On the right-hand panel of 
Fig.~\ref{fig:beta-tau2}, this residual $A$ dependence of $\tau_1/\beta_1$ has 
been fitted by the power law $aA^b$ with $b\sim -1/3$, again omitting the iron 
data due to their unsystematic value of $\epsilon$. Thus, $\tau_1/\beta_1$ will
approach the real axis for large mass numbers. Therefore, one can conclude that
the modeled pairing phase transition is a true phase transition in the 
generalized Ehrenfest sense, provided that one investigates this phase 
transition with respect to the relevant size of the system, which is the mass 
number $A$ rather than the size of the model space $N$. As a final remark, it 
is interesting to notice that the odd-even difference of $\beta_1$ and 
$\tau_1/\beta_1$ becomes smaller with increasing mass number. Thus, in the 
thermodynamical limit, it will not matter if the system consists of an even or 
an odd number of particles, as it is expected.

\section{Conclusions}

We have developed a model which allows the investigation of phase transitions 
in a small system of two or more particles. The anticipated phase transition 
from a paired to an unpaired system in rare earth nuclei could be produced. The
order of the modeled pairing phase transition could be determined for even 
systems to be of second order and for odd systems to be of higher order. A 
second phase transition, which is of first order, is inherent in the model and 
connects the unpaired and a quasi-classical phase. The latter phase is 
characterized by the dilution of particles in phase space and might therefore 
be encountered in systems where the phase space is much larger in size (in 
terms of the number of realizations of thermal excitations) than the number of 
particles present. This second phase transition is a true phase transition and 
might be realized in nature by some physical system. In the case of modeling 
heavy nuclei with particle numbers much larger than the model space, however, 
it reflects merely the fact that the model space is still too small in order to
describe those nuclei adequately at high temperatures (within the limits 
discussed in \footnotemark[4]). It has been shown that the pairing phase 
transition and the signal of the exhaustion of the finite model space have to 
be carefully separated in temperature in order to obtain a clear signature of 
the pairing phase transition in atomic nuclei. The presented model is able to 
take into account a sufficient number of single-particle levels and nucleons to
allow a good separation and to display conclusive evidence of the pairing phase
transition in heavy mid-shell nuclei.

\acknowledgements

Part of this work was performed under the auspices of the U.S. Department of 
Energy by the University of California, Lawrence Livermore National Laboratory 
under Contract No.\ W-7405-ENG-48. Financial support from the Norwegian 
Research Council (NFR) is gratefully acknowledged. We thank David Dean and Paul
Garrett for interesting discussions and Emel Tavukcu for carrying out some 
preliminary calculations for this work.

\end{multicols}

\clearpage

\begin{table}\centering
\caption{The model parameters have been calculated by means of 
$\epsilon=30$~MeV/$A$, $\Delta=12$~MeV/$\sqrt{A}$, and 
$A_{\mathrm{rig}}=34$~MeV~$A^{-5/3}$ with the exception of $\epsilon$ for 
$^{58}$Fe (see text). All values are rounded to integer numbers. The values for
$\hbar\omega_{\mathrm{vib}}$ have been taken from spectroscopic data of 
Ref.~[36].
The value of $r=0.56$ is suggested for the mass 190 region by the calculations 
of Ref.~[37] 
and adopted for all four nuclei in this work.}
\label{tab:parameters}
\begin{tabular}{cccccc}
Nucleus&$\epsilon$ [keV]&$\Delta$ [keV]&$A_{\mathrm{rig}}$ [keV]&$\hbar
\omega_{\mathrm{vib}}$ [MeV]&$r$\\\hline
$^{58}$Fe&800&1576&39&2.0&0.56\\
$^{106}$Pd&283&1166&14&1.4&0.56\\
$^{162}$Dy&185&943&7&0.9&0.56\\
$^{234}$U&128&784&4&0.8&0.56\\
\end{tabular}
\end{table}

\clearpage

\clearpage

\clearpage

\clearpage

\clearpage

\clearpage

\clearpage

\clearpage

\clearpage

\end{document}